# Machine Learning for the Zwicky Transient Facility

Ashish Mahabal[1,2], Umaa Rebbapragada[3], Richard Walters[4], Frank J. Masci[5], Nadejda Blagorodnova[1], Jan van Roestel[6], Quan-Zhi Ye (葉泉志)[1,5], Rahul Biswas[7], Kevin Burdge[1], Chan-Kao Chang (章展誥)[8], Dmitry A. Duev[1], V. Zach Golkhou[9,10,30], Adam A. Miller[11,12], Jakob Nordin[13], Charlotte Ward[14], Scott Adams[1], Eric C. Bellm[9], Doug Branton[15], Brian Bue[3], Chris Cannella[1], Andrew Connolly[16], Richard Dekany[4], Ulrich Feindt[7], Tiara Hung[14], Lucy Fortson[17,18], Sara Frederick[14], C. Fremling[1], Suvi Gezari[14], Matthew Graham[19], Steven Groom[5], Mansi M. Kasliwal[19], Shrinivas Kulkarni[19], Thomas Kupfer[19,20,21], Hsing Wen Lin (林省文)[8,22], Chris Lintott[17,23], Ragnhild Lunnan[24], John Parejko[16], Thomas A. Prince[19], Reed Riddle[4], Ben Rusholme[5], Nicholas Saunders[16], Nima Sedaghat[25], David L. Shupe[5], Leo P. Singer[26,27], Maayane T. Soumagnac[28], Paula Szkody[9], Yutaro Tachibana (優太朗橘)[1,29], Kushal Tirumala[19], Sjoert van Velzen[14], and Darryl Wright[17,18]

[1] Division of Physics, Mathematics, and Astronomy, California Institute of Technology, Pasadena, CA 91125, USA; aam@astro.caltech.edu
[2] Center for Data Driven Discovery, California Institute of Technology, Pasadena, CA 91125, USA
[3] Jet Propulsion Laboratory, California Institute of Technology, Pasadena, CA 91125, USA
[4] Caltech Optical Observatories, California Institute of Technology, Pasadena, CA 91125, USA
[5] Infrared Processing and Analysis Center, California Institute of Technology, MS 100-22, Pasadena, CA 91125, USA
[6] Department of Astrophysics/IMAPP, Radboud University Nijmegen, P.O.Box 9010, 6500 GL, Nijmegen, The Netherlands
[7] The Oskar Klein Centre, Department of Physics, Stockholm University, AlbaNova, SE-106 91 Stockholm, Sweden
[8] Institute of Astronomy, National Central University, 32001, Taiwan
[9] Department of Astronomy, University of Washington, Seattle, WA 98195, USA
[10] The eScience Institute, University of Washington, Seattle, WA 98195, USA
[11] Center for Interdisciplinary Exploration and Research in Astrophysics and Department of Physics and Astronomy, Northwestern University, 2145 Sheridan Road, Evanston, IL 60208, USA
[12] The Adler Planetarium, Chicago, IL 60605, USA
[13] Institute of Physics, Humboldt-Universität zu Berlin, Newtonstr. 15, 12489 Berlin, Germany
[14] Department of Astronomy, University of Maryland, College Park, MD 20742, USA
[15] Space Telescope Science Institute, 3700 San Martin Drive, Baltimore, MD, 21218, USA
[16] DIRAC Institute, Department of Astronomy, University of Washington, 3910 15th Avenue NE, Seattle, WA 98195, USA
[17] Zooniverse, Oxford Astrophysics, Denys Wilkinson Building, Keble Road, Oxford, OX1 3RH, UK
[18] School of Physics and Astronomy, 116 Church Street SE, University of Minnesota, Minneapolis, MN 55455, USA
[19] Department of Astronomy, California Institute of Technology, Pasadena, CA 91125, USA
[20] Kavli Institute for Theoretical Physics, University of California, Santa Barbara, CA 93106, USA
[21] Department of Physics, University of California, Santa Barbara, CA 93106, USA
[22] Department of Physics, University of Michigan, Ann Arbor, MI 48109, USA
[23] Department of Physics, The University of Oxford, Keble Road, Oxford, OX1 3RH, UK
[24] The Oskar Klein Centre & Department of Astronomy, Stockholm University, AlbaNova, SE-106 91 Stockholm, Sweden
[25] Department of Computer Science, University of Freiburg, Georges-Koehler-Allee 052, 79110 Freiburg, Germany
[26] Astrophysics Science Division, NASA Goddard Space Flight Center, MC 661, Greenbelt, MD 20771, USA
[27] Joint Space-Science Institute, University of Maryland, College Park, MD 20742, USA
[28] Department of Particle Physics and Astrophysics, Weizmann Institute of Science 234 Herzl Street, Rehovot, 76100, Israel
[29] Department of Physics, Tokyo Institute of Technology, 2-12-1 Ookayama, Meguro-ku, Tokyo 152-8551, Japan
Received 2018 September 4; accepted 2018 November 26; published 2019 January 31

## Abstract

The Zwicky Transient Facility is a large optical survey in multiple filters producing hundreds of thousands of transient alerts per night. We describe here various machine learning (ML) implementations and plans to make the maximal use of the large data set by taking advantage of the temporal nature of the data, and further combining it with other data sets. We start with the initial steps of separating bogus candidates from real ones, separating stars and galaxies, and go on to the classification of real objects into various classes. Besides the usual methods (e.g., based on features extracted from light curves) we also describe early plans for alternate methods including the use of domain adaptation, and deep learning. In a similar fashion we describe efforts to detect fast moving asteroids.

---

[30] Moore/Sloan Foundation, WRF Innovation in Data Science, and DIRAC Fellow.

We also describe the use of the Zooniverse platform for helping with classifications through the creation of training samples, and active learning. Finally we mention the synergistic aspects of ZTF and LSST from the ML perspective.

*Online material:* color figures

## 1. Introduction

For the last couple of decades sky surveys covering increasingly larger areas and depth have provided a wealth of information. A few examples of such surveys are Digitized Palomar Observatory Sky Survey (DPOSS; Weir 1995), Palomar-QUEST (PQ; Djorgovski et al. 2008), Palomar Transient Factory (PTF; Law et al. 2009), Catalina Real-time Transient Survey (CRTS; Drake et al. 2009; Mahabal et al. 2011), Pan-STARRS (Kaiser 2004), ASASSN (Shappee et al. 2014), ATLAS (Tonry et al. 2018), *Gaia* (Gaia Collaboration et al. 2016), etc. They have provided the ability to observe changes in hundreds of millions of sources, leading to an understanding of large families of sources. More importantly, combined with faster computers, and availability of compatible historic data, the surveys now routinely enable real-time follow-up of rapidly fading transient sources and of interesting variable sources.

*Gaia* is revolutionizing Galactic astronomy, especially of stars with astrometrically unsurpassed observations with ∼70 observations over five years down to $r \approx 20$ mag, while deeper surveys like the LSST (Ivezić et al. 2008) loom nearby, capable of reaching $r \approx 24$ mag in a single exposure, providing about 1000 observations for the observable sky over 10 years. The Zwicky Transient Facility (ZTF) described here uses the Palomar 48-inch Schmidt, and hits the sweet spot between these in terms of depths, reaching $r \approx 21$ mag, accompanied by an extremely high cadence facilitated by a large 47 $\deg^2$ field of view effectively maximizing the volume of sky where astrophysical sources of brightnesses suitable for spectroscopic follow-up are likely to be found (Bellm 2016). More details about the survey can be found in Bellm et al. (2019). In just three years, ZTF is projected to yield about thousand observations of the sky observable from the Palomar observatory, produce hundreds of thousands of alerts every clear night that could be spectroscopically followed up by 1–5$m$ diameter telescopes. Such a discovery and follow-up program will create a large repository of transient and variable sources, that will be useful for analysis of the upcoming deeper, largely photometric surveys like LSST. Detailed science drivers for ZTF are described in Graham et al. (2019).

The rich ZTF data set brings along technologically interesting problems. While we would gain tremendous knowledge by taking spectra of all the transient and variable sources, spectroscopic resources are limited, and moreover, a large fraction of the variable sources belong to classes of objects that are relatively well understood. Thus, the biggest needs are to select (a) a subset of objects for follow-up that will best increase our understanding of the different families, and (b) include rapidly fading objects that provide a small temporal window for follow-up. Given the large number of nightly sources, this necessitates automated early probabilistic characterization of sources. With surveys like ZTF we are starting to reach alert volumes that are beyond vetting on an individual basis. One could employ an effort heavy on citizen science at the cost of loss in fidelity, or pure machine learning that can lead to a loss in accuracy if carried out blindly. We describe here the combination approach we plan to take, bordering on active learning with human involvement.

Overall, especially given that this is a survey with 16 new large CCDs and a new camera, we need to worry about two separate classification problems: (a) real/bogus (RB) to separate astrophysically genuine objects from artifacts, and (b) to then separate into different classes from among the astrophysical sources. Asteroids are marked as such through cross-matching with known catalogs and dealt with separately. A software filter applied to real objects by different science working groups requires two detections separated by a suitable interval at the same location. This eliminates asteroids not already known (perhaps at the cost of very rapid transients). When it comes to classification of non-moving astrophysical sources, we have to contend with varying colors, different rise and fall times, and population densities as a function of several variables, making the problem more challenging. We describe various machine learning methods as well as the use of external brokers and additional filters for early characterization, some in operation, and some under development. Section 2 has the overall data flow; Section 3 the real/bogus processing; Section 4 the methods employed for solar system objects; Section 5 the planned classification efforts for stellar and extragalactic objects; and Section 6 the description of some synergistic brokering and machine learning efforts.

This paper is complemented by a few other papers that go into greater detail on many aspects of ZTF: technical specifications, ZTF camera, and survey design (Bellm et al. 2019), the observing system and the instrumentation (Dekany et al. 2018), the science data system (Masci et al. 2019), the alert system (Patterson et al. 2019), the GROWTH marshal (Kasliwal et al. 2019), the star-galaxy separation (Tachibana et al. 2018), and the overall science objectives of ZTF (Graham et al. 2019).

## 2. Data Flow, Preprocessing, and Computational Setup

The ZTF camera consists of 16 CCDs, each with four readout channels of 3k × 3k pixels. Approximately 700 science exposures are observed on an average clear night, yielding about 1 TB of uncompressed data. Data are processed and stored at IPAC along with metadata and resulting products. The processing includes image differencing i.e., the subtraction of a reference image from the science image. This is done using the ZOGY algorithm (Zackay et al. 2016). A list of sources is extracted from this image. The subtraction is also done in the reverse direction to look for fading sources relative to the reference image level. For objects found in this reverse subtraction, the *isdiffpos* flag is set to zero. Such metadata are useful for users wanting to select a subset of objects from the public stream of alerts described below. The number of unfiltered $5\sigma$ alerts, depending on sky position and availability of reference images, varies from $\sim 10^5$ to potentially $3 \times 10^6$. The alerts are made available as Avro[31] packets using the Kafka[32] system. Each packet consists of an *objectid*, source-specific features based on the difference image; the metadata specific to science; the reference; the difference images; the count of previous detections; the history of up to 30 days; the nearest solar system objects; the three closest PS1 objects within a certain radius; and the $63 \times 63$ pixel$^2$ cutouts for science; reference; difference images; and the real/bogus score (see Section 3) from the latest deployed model. More details can be found in Masci et al. (2019) and Patterson et al. (2019).

To eliminate obvious image artifacts (false positives) in the raw candidate stream from difference images, the following initial cuts are applied:

1. Detection signal to noise (S/N): S/N > 5; this S/N is from the ZOGY point source match-filtered image;
2. Photometric S/N > 5; based on an 8-pixel diameter circular aperture;
3. Detection is >10 pixels from an image edge;
4. Source elongation (A/B from fitted elliptical profile) $\leqslant 2$;
5. Ratio of fluxes, R, satisfying: $0 < R \leqslant 1.5$ where R = flux in 8-pixel diameter aperture/flux in 18-pixel diameter aperture;
6. Number of negative pixels in a $5 \times 5$ pixel area $\leqslant 13$;
7. Number of bad pixels in a $5 \times 5$ pixel area $\leqslant 7$; and
8. Absolute difference between PSF and aperture photometry $\leqslant 1$ mag.

### 2.1. Computing Setup

To support time domain astronomy with ZTF in general and the machine learning activities in particular, we have built a dedicated "database machine," a server running Red Hat Enterprise Linux with 1 TB of DDR4 memory, two 10-core Intel Xeon CPUs, and 50 TB of storage in a RAID6 configuration with a dedicated RAID card. We chose the *MongoDB*[33] NoSQL database as the database engine with built-in *GeoJSON* support with a range of supported geometries and 2D indexes on a sphere that enable extremely fast positional queries: a typical cone search query on a collection —an analog of a table in traditional SQL databases—whose index fits into memory takes about 20–50 μs.

The full ZTF alert stream is saved to a dedicated collection. Besides the ZTF light curve collection, over 20 catalogs (including Gaia DR2 and Pan-STARRS) are currently available for querying and cross-matching; more catalogs will be ingested into the database in the near future. We have developed an API that allows access to the service both programmatically with a *python* client and through a web-based interface.[34] It is powered by a multi-threaded *Flask*[35]-based back-end running behind an *nginx*[36] reverse proxy server. The API is *socket.io*[37]-based and is capable of handling tens of thousands of simultaneous connections. A distributed queue system built with *redis*[38] and $python - rq$[39] is used to execute queries. The communication with the server is encrypted and JSON Web Tokens[40] are used for user authentication. To simplify maintenance, monitoring, and deployment the system is containerized using *Docker*.[41]

### 2.2. Star-galaxy Classification

To support real-time classification, and the Galactic variability survey, we have constructed a new model to identify unresolved point sources detected in the Pan-STARRS1 (PS1; Chambers et al. 2016) survey. The full details of the random forest model used to perform this classification are given in Tachibana et al. (2018). Here we describe advantages thereof for downstream machine learning. Briefly, the PS1 catalogs provide many advantages relative to the PTF catalogs that we previously used to identify point sources (Miller et al. 2017), including (i) better sky coverage (a dead CCD in the PTF camera resulted in $\sim 1/12$ of the northern sky not being covered by PTF); (ii) deeper and more uniform imaging (the PTF cadence varied from field to field leading to significant discrepancies in the final stack depth); and (iii) the use of five filters (the PTF point source catalog included only $R_{\rm PTF}$ detections). The features for the final PS1 model utilize signal-to-noise weighted mean values for the flux and shape measurements provided in the PS1 database. The model is trained using morphological classifications of ∼48,000 sources detected by the *Hubble Space Telescope* COSMOS survey (Leauthaud et al. 2007). We adopt the same figure of merit (FoM) as Miller et al. (2017), namely to maximize the true positive rate (TPR) of stellar classifications at a fixed false positive rate (FPR) of 0.005, and achieve a FoM ≈ 0.7, which is demonstrated to outperform the SDSS, PS1, and PTF photometric classifiers (see Tachibana et al. 2018 for further details). Moving forward, we plan to update the point source catalog by incorporating stars from the *GAIA* catalog, which we will further supplement with PS1 data release 2.

---

[31] http://avro.apache.org/
[32] https://kafka.apache.org
[33] https://www.mongodb.com/
[34] https://github.com/dmitryduev/broker/
[35] http://flask.pocoo.org/
[36] https://nginx.org/
[37] https://socket.io/
[38] https://redis.io/
[39] http://python-rq.org/
[40] https://jwt.io/
[41] https://docker.com/

## 3. Real/Bogus Separation

Detection of transients can be done in the catalog domain (e.g., CRTS survey; Drake et al. 2009), or in the image domain (e.g., PTF survey; Law et al. 2009) where one has to contend with matching PSF, depth, and other temporal characteristics. This can lead to a large number of artifacts (*bogus* events or false positives) compared to genuine astrophysical (*real*) sources. This led to the introduction of a Real/Bogus classifier (RB) that scores individual sources on a scale of 0.0 (bogus) to 1.0 (real). RB classifiers were introduced by Bailey et al. (2007) for the Nearby Supernova Factory (Aldering et al. 2002), and have been adopted by other time domain surveys including the PTF (Bloom et al. 2008) and the Intermediate Palomar Transient Factory (iPTF; Brink et al. 2012; Wozniak et al. 2013; Rebbapragada et al. 2015); the Dark Energy Survey (Goldstein et al. 2015); and Pan-STARRS (Wright et al. 2015).

Currently employed RB classifiers are built using supervised machine learning algorithms that take as input a set of candidate sources that have been labeled as real or bogus. The candidates themselves are represented through a series of measurements and observables, generically called *features*. The features are extracted from science and subtracted image cutouts centered on the candidate, and supplemented with other measurements taken from science, subtracted and reference images (see Table 1 for a full list of features).

The challenge of RB classification is the construction of a training set that is representative of nightly data across filters, sky location, and in the case of multi-CCD surveys like ZTF, possible variations between CCDs, as well as cross-talk. The bogus to real ratio in PTF was an estimated 1000:1 (Bloom et al. 2008), though Brink et al. (2012) discussed the difficulty in calculating the true ratio due to sampling bias inherent in labeled data sets. At the end of the survey, PTF and iPTF RB systems were trained with tens of thousands of candidates. ZTF uses a different camera and 16 new 6k × 6k CCDs together having 64 amplifiers leading to somewhat different artifacts. A new subtraction scheme (using ZOGY; Zackay et al. 2016) has reduced the number of artifacts drastically, but also altered the nature of artifacts that are currently seen. As a result, we have not been able to leverage the iPTF work for training though this is the subject of future work using domain adaptation.

The first ZTF RB for point source candidates was deployed with 1620 training examples collected from science validation imaging through 2108 January 10. 1316 were labeled real (81.6%) and 304 (18.4%) were labeled bogus. The labeling was done by members of the ZTF collaboration on the Zooniverse Citizen Science platform[42], as first developed for galaxy morphology (Lintott et al. 2008). Here, members are shown $63 \times 63$ pixel$^2$ thumbnails centered on the candidates from the science, reference and difference images, along with metadata about the source (see Table 2). Details about the Zooniverse setup are provided in Section 3.1.2. All available labeled data are used to train a random forest classifier. We use the ExtraTrees classifier (Geurts et al. 2006), a variant of the random forest classifier (Breiman 2001), as implemented in scikit-learn (Pedregosa et al. 2011). We train 300 trees and consider the square root of the number of features when determining node splits. We measure the performance of the random forest via accuracy, false positive and negative rates over ten-fold cross validation, and also by examining score distributions on independent tests of candidates. We name classifiers according to the versions of the training samples, feature set, and pipeline software. For example, the initial RB version was referred as t1_f1_c1 (*t* for training sample, *f* for feature set, and *c* for software version). The false positive rate (FPR), false negative rate (FNR), and the accuracy (ACC) of classifier t1_f1_c1 were 30.7%, 3.8%, and 91.2% respectively. We also calculate the FNR at a desired 1% FPR, which for t1_f1_c1 was 36.4%. In comparison, the final classifier for iPTF had a training set of 10,000 examples and a FNR of 5.7% at 1% FPR.

To improve performance, we retrain and deploy new classifiers as more labeled data are collected. A second source of labeled data is the GROWTH marshal (Kasliwal et al. 2019), where team members label bogus and real objects as they scan nightly alerts, often with specific filtering tuned for their science program. Details about the marshal setup are provided in Section 3.1.1.

Classifier t8_f5_c3 was trained on labeled objects through 2108 July 5 and tested on objects from 2018 July 6 to August 3. This version (t12_f5_c3) had an associated FPR, FNR, ACC, and FNR at 1% FPR of 14.9%, 7.2%, 89.0%, and 33.6%, respectively. Figures 1 and 2 show the difference in the cumulative score distributions on independent test sets of real and bogus objects for t1_f1_c1 and t8_f5_c3, with t8_f5_c3 scoring bogus objects lower and real objects higher.

Classifiers since then have shown smaller, incremental improvements for specific subsets; for instance for low Galactic latitude crowded fields. We continue to explore improvements to the RB training process, tracking and removing label contamination. Greater details and performance about the RB training process will be presented in a forthcoming paper.

---

[42] https://www.zooniverse.org/

Table 1
Features Used by the Random Forest Classifier. In the Feature Names, *sci* Refers to Science Images, *ref* to Reference Images, and *diff* to Difference Images

| | |
|---|---|
| **Image Level Features** | |
| Limiting magnitude | Expected 5σ magnitude limit of the *sci* and *ref* images after gain and background matching (and resampling for the reference image). Expected 5σ magnitude limit of diff image. |
| Flux ratio | Median flux ratio ($Flux_{sci}/Flux_{ref}$) across matched sources. |
| Sci image | Electronic gain (small fluctuations are observed), saturation level (after gain-matching), modal background level, and robust sigma/pixel after gain and background matching. |
| Ref image | Saturation level (after gain-matching) and resampling, modal background level, and robust sigma/pixel after gain, background matching, and resampling. |
| Diff image | Robust sigma/pixel, number of bad pixels before and after PSF matching; percentage of pixels that are bad or unusable; effective FWHM in *diff* image; the average of squared *diff* image pixel values before and after PSF matching; percentage of changed *diff* image pixels values before and after PSF matching; and status of image differencing (0 = bad, 1 = good). |
| Positive versus negative diff | Median background level in positive and negative *diff* image, and number of candidates extracted from the positive and negative *diff* image before and after internal filtering. |
| Signal-to-noise ratio (S/N) in *diff* image | Number of noise pixels in diff image, peak-pixel S/N in detection image optimized for point source detection, and ratio of the mean square pixel value in the subtraction image to its spatial variance following PSF matching. |
| Seeing | Seeing of *sci* and *ref* images, and ratio of FWHM to average FWHM on the *sci* image. |
| **Candidate Features** | |
| PSF photometry | Magnitude and 1σ uncertainty from PSF fit; chi of candidate |
| Aperture photometry | Magnitude and 1σ uncertainty from "big" aperture photometry |
| Candidate measurements | Local sky background level; FWHM of local Gaussian profile; magnitude difference of PSF and aperture photometry; magnitude difference of PSF photometry and limiting magnitude; and minimum distance to edge. |
| Nearest *ref* source | Distance, magnitude, 1σ uncertainty, chi and sharpness of nearest reference image extraction. |
| Nearest solar system object | Distance and magnitude of nearest solar system object |
| Shape | Windowed rms along major; minor axis of source profile; ratios of the major and minor axes to the FWHM; and elongation of the candidate. |
| Negative/bad pixels | Number of negative pixels in a 7 × 7 box, number of bad pixels in a 7 × 7 box, ratio of sum of pixel values to sum of absolute values in source-stamp cutout. |

Table 2
Metadata Provided for Zooniverse Classifications

| Parameter | Description |
|---|---|
| magpsf | Magnitude of object in science image |
| sigmapsf | Error of magnitude in science image |
| classtar | Sextractor likelihood of stellarity (1 = highest, 0 = lowest) |
| ssdistnr | Distance to nearest solar system object |
| sgscore | Star galaxy score |
| isdiffpos | The object in the difference is a positive subtraction |

### 3.1. Candidate Labeling for RB Classifier Training

#### 3.1.1. Labeling Using the Transient Marshal

Different science groups filter the alerts based on their science case of interest using the GROWTH marshal (Kasliwal et al. 2019), which allows the configuration of custom filters. For example, the group interested in tidal disruption events (TDEs) will filter candidates that have a known star within a pixel, whereas the group interested in Galactic variables will not. Groups interested in rapidly evolving candidates construct their filters to only accept candidates whose magnitudes rise more quickly than some specified rate. To avoid accepting asteroid detections, most programs have filters that only accept candidates having multiple detections across a set period of time (often 15 to 30 minutes). This common asteroid rejection criteria results in most labels from the GROWTH marshal

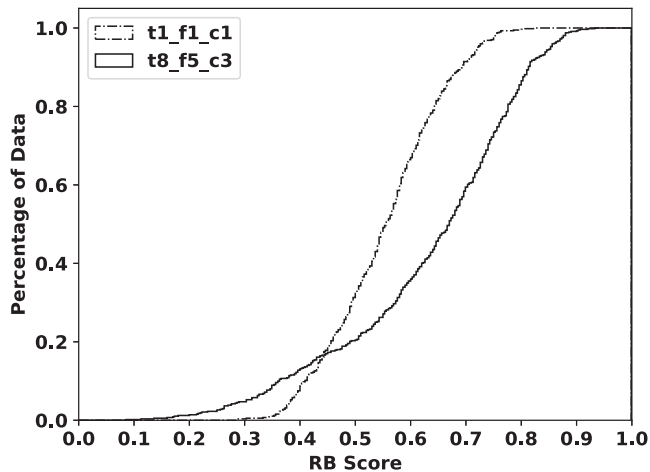

**Figure 1.** Comparison of cumulative distributions generated by the first RB version (t1_f1_c1; solid) with a recent (t8_f5_c3; dashed) one on a test set of 572 vetted reals spanning 2018 May 8 through 24. It can be seen that ∼20% reals have an RB score below 0.5 compared with ∼40% earlier. This should keep getting better as more diverse reals get included in the trainings set.

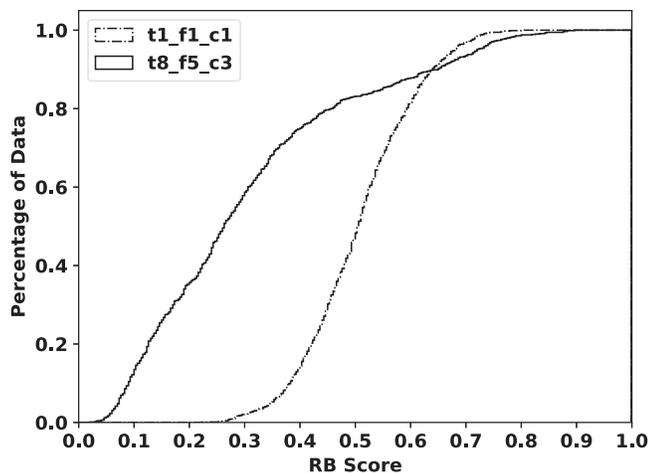

**Figure 2.** Comparison of cumulative distributions generated by the first RB version (t1_f1_c1) with a recent one (t8_f5_c3) on a test set of 920 vetted bogus candidates spanning 2018 May 8 through 24. Earlier ∼50% bogus objects had an RB score of over 50%. That has now dropped to ∼25%, and getting better. Training now focuses on the volume where there is greatest overlap of real and bogus objects.

being for candidates that are persistent through multiple observations.

As each science team reviews candidates, members label objects as real or bogus, as well as selecting a subset of real objects for follow-up. Labeling through the GROWTH marshal can result in a biased training set because only candidates for science programs are reviewed. In future work, we will address the problem of biased sample selection through the application of active learning (Settles 2012) to examples reviewed through the Zooniverse web interface.

### 3.1.2. Labeling Using Zooniverse

The Zooniverse citizen-science platform is a robust platform for data visualization and classification/labeling. It provides the necessary tools for data upload, classifications, and downloading of reports. A Zooniverse project requires a subject set (e.g., set of candidates along with their metadata) and a workflow (a sequence of specific tasks for the end user to perform). We currently use it to supplement the set of real and bogus labels that come from the marshal by running targeted campaigns whereby we switch in different data sets for specific number of days, and get classifications on that set. The set of campaigns running early on include those targeting specific CCDs so that we can understand the distribution of artifacts for each CCD.

The current workflow consists of just one task: classify if an alert is real or bogus based on the science, reference, and difference images, associated metadata, and a PS1 cutout that is provided for its greater depth (see Figure 3). A skip option is provided for ambiguous cases (see Figure 4). Figure 5 shows all the panels as seen by the citizen scientists.

During the commissioning period there have been several thousand classifications including ∼6000 real, ∼1800 bogus, and ∼1200 skipped ones. Users labeled subsets of objects from RB intervals like [0.0, 0.2], [0.8, 1.0], and [0.4, 0.6]. Objects in the first two interval sets catch obvious mislabeled examples, while objects in the last set are those on which the current classifier is uncertain. The labeling of these intermediate objects can greatly clarify the decision boundaries. We have involved only ZTF members in Zooniverse classification so far.

We will soon be involving a wider set of volunteers on public ZTF data. The Zooniverse infrastructure also allows for the integration of active learning techniques (Settles 2012), where subjects ranked by some confidence heuristic can be prioritized for human screening. We plan to integrate active learning into our Zooniverse project in future work.

### 3.2. Characterization of Real/bogus

It is crucial to understand the characteristics of ZTF detections to understand the parameter space occupied by real and bogus detections. That can be done using a list of candidates vetted by ZTF users, and by comparing sources discovered by surveys independent of the ZTF alerting system. A subsequent forced analysis at such locations can reveal possible reasons for the lack of completeness.

It is important to understand the sources of contamination by analyzing sources initially marked as positive and/or with high RB scores that subsequently turn out to be not alert-worthy. This can lead to a purer sample and an increase in throughput.

Our experiment consisted in separating the parameter region where real and bogus detections existed. Using two lists with objects visually saved by users as real and bogus respectively, we obtained a list of ZTF alerts associated with each object (since the start of the alert stream). We then analyzed the most representative features, e.g., FWHM, magnitude, signal-to-noise (S/N), for those alerts.

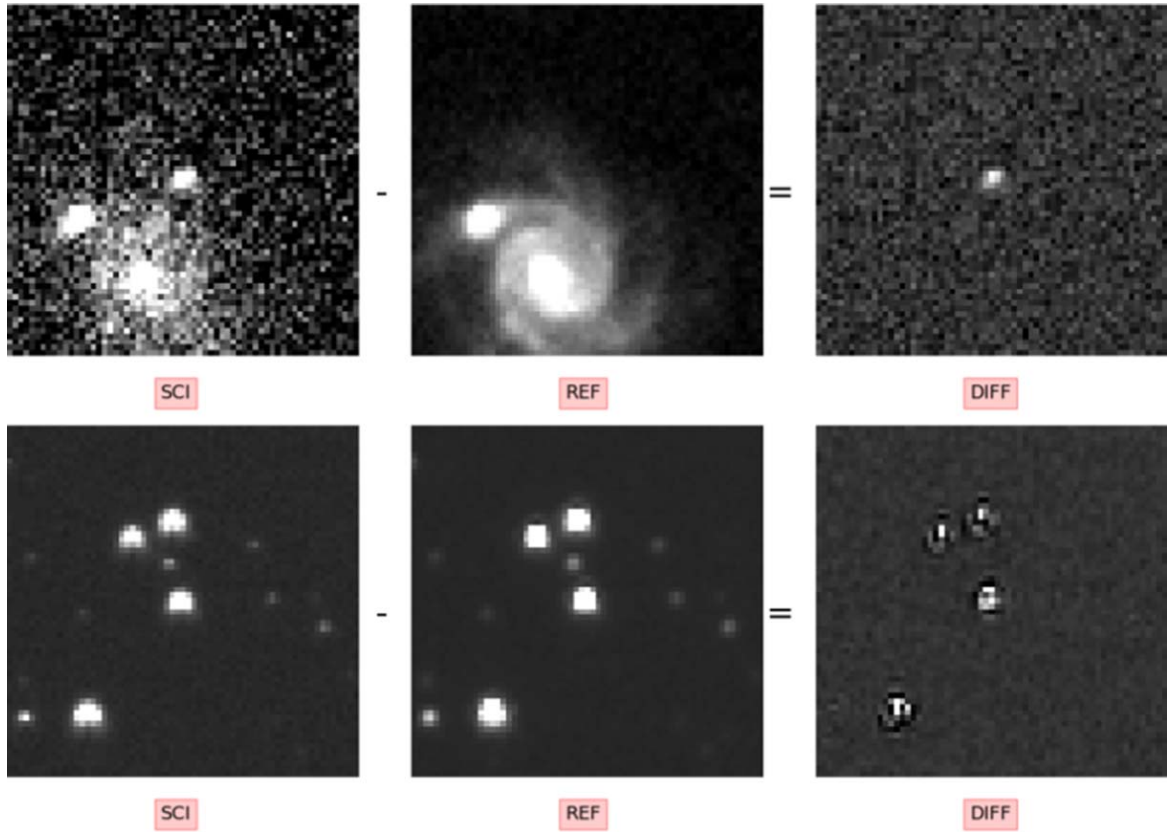

**Figure 3.** Zooniverse images for a real (top panel) and a bogus (bottom panel) transient. In each panel, from left to right we show the science, reference, and difference images.
(A color version of this figure is available in the online journal.)

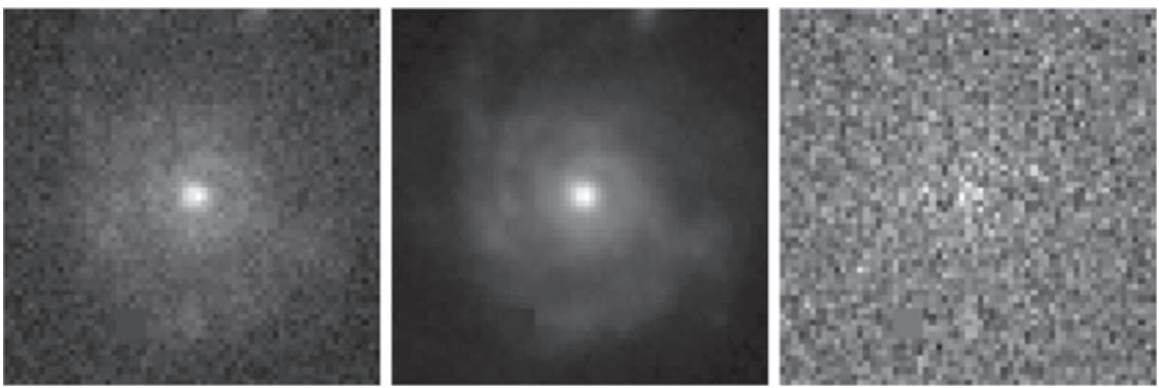

**Figure 4.** Subtraction artifact at the center of a galaxy. As in Figure 3, the image on the left is the science image, followed by the reference image, and finally the difference image.

For example, real objects would be expected to have shapes comparable to the image's PSF, with a FWHM corresponding to the average seeing in the image. Bogus objects, which pass other tests, on the contrary, generally tend to be fuzzier, with larger FWHM. Figure 6 shows an example of the density distributions for the two groups. Real detections are, as expected, mostly centered around lower values of FWHM. Because they are point-like sources, their measured aperture magnitude and PSF magnitude are similar. Alerts associated with bogus detections show a larger spread in their FWHM, reaching unreasonably high values, and consequently having fainter PSF magnitudes than aperture magnitudes. The ZTF camera employs a number of dedicated focusing and guiding CCDs commissioned in 2018 Q2/Q3. All of these observations are from before the focus CCDs were in place, so the foci for many parameters should get tighter and help further with the real/bogus separation.

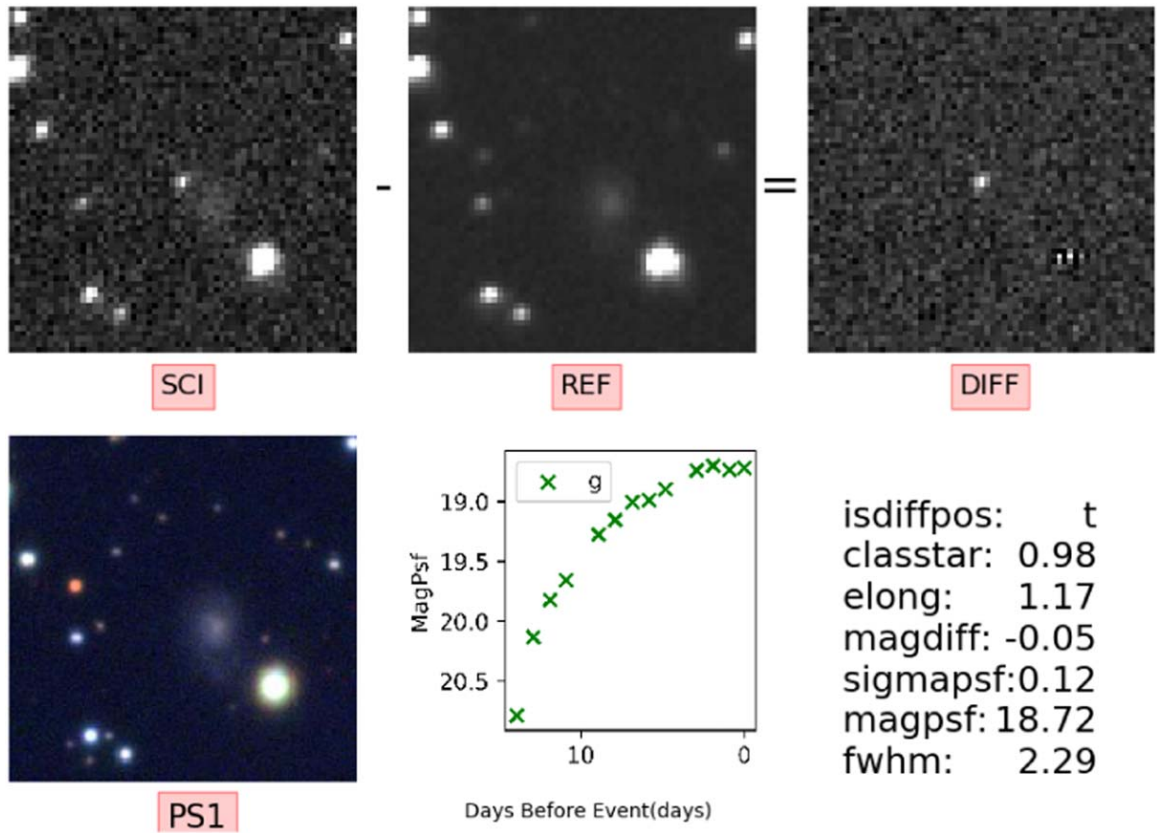

**Figure 5.** Panel layout for the Zooniverse MSIP project. In addition to the science, reference, and difference images in the top row, we see, from the left, the following in the bottom row: an archival image from PS1, light curve within the preceding 30 days, and some metadata useful for classification.
(A color version of this figure is available in the online journal.)

We carried out a similar analysis for other features, such as the elongation of the detection, its S/N, and the number of bad and negative pixels in the detection. We have implemented the following cuts, in addition to S/N > 5 requirement. Alerts are excluded if any of these conditions are met:

1. PSF mag > 23.5,
2. Number of bad pixels > 4,
3. FWHM > 7,
4. Elongation > 1.6,
5. aperture mag - PSF mag > 0.4, and
6. aperture mag - PSF mag < −0.75.

We will also be watching an envelope volume around these limits to collect bogus objects that may mimic the reals and closely investigate them. The cuts are generous so as to not exclude real objects. These procedures, aided by standard visualization, have helped us cut down the initial bogus candidates by well over 50%, with an estimated loss of a couple percent for the reals (based so far on small number statistics). We are still in the early days, and the data paucity implies scope for improvement in the models, a process that will continue as we gather more data.

### 3.2.1. Verifications Using External Transient Surveys

We run a verification of our classification system using the transient discoveries reported to the Transient Name Server (TNS).[43] We start with non-ZTF TNS objects brighter than 19.5 magnitudes (in any filter), with decl. > −20°, and between two specific dates (2108 June 2 through August 14). We cross-match these with ZTF alerts within 3 arcsec. We select only ZTF alerts generated up to one week before the public announcement and two weeks after.

---
[43] https://wis-tns.weizmann.ac.il/

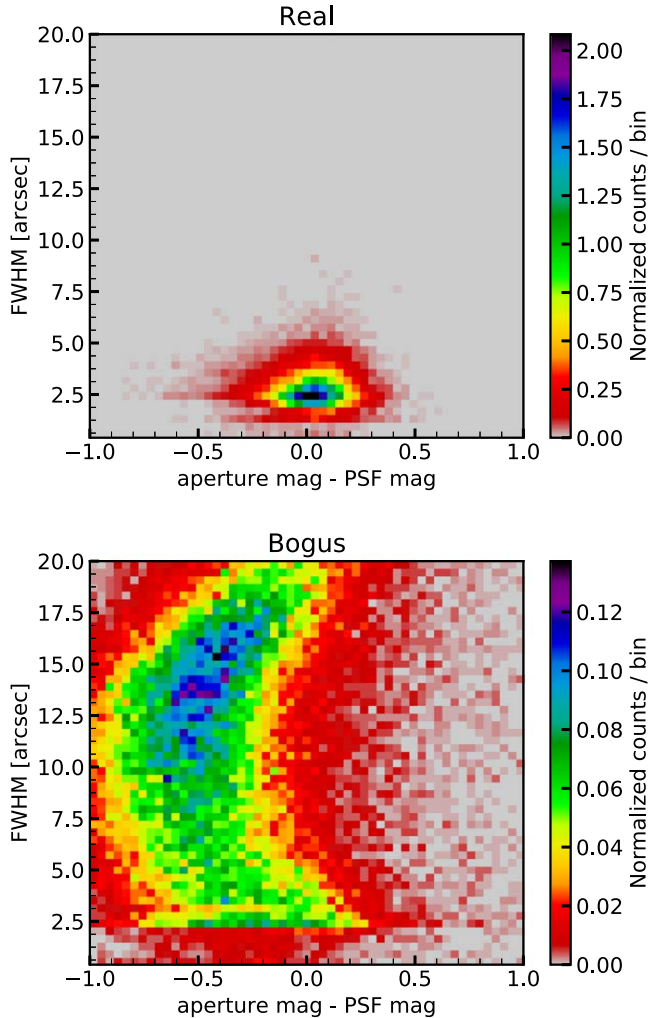

**Figure 6.** Two-dimensional histogram for the FWHM vs. the magnitude difference (aperture magnitude minus the PSF magnitude) for alerts matched to real transients (top) and matched to bogus detections (bottom). We used 21,230 alerts associated with real objects and 64,875 associated with bogus detections. Here and in Figure 7, all points from light curves of the real and bogus candidates have been used. This is in contrast to the RB training which uses only the discovery points. For the purpose of this analysis, the bogus objects include variable stars (that may initially have been marked as bogus due to saturation, for instance), hence many of them have multiple points as well.
(A color version of this figure is available in the online journal.)

From this set, we plot the distribution of the real/bogus score for real TNS transients with ZTF alerts (see Figure 7). The figure also compares the PSF magnitude of each ZTF detection with its real/bogus score and its S/N.

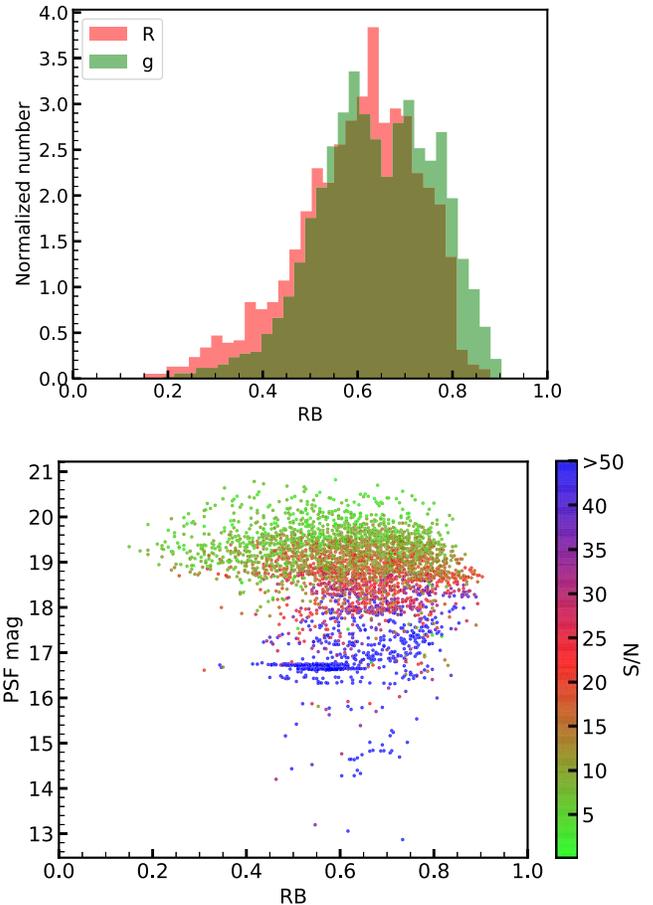

**Figure 7.** Top: RB score distribution in each band for ZTF alerts matched to TNS sources with discovery dates between UTC 2018 June 2 to August 14. Bottom: Scatter plot of the PSF magnitudes of the transients vs. their RB scores. The color scale indicates the S/N of the residual.
(A color version of this figure is available in the online journal.)

The RB score distribution for the alerts shows a group with high scores, around 0.7 for the $r$ band and a bimodal distribution from the $g$ band, centered between 0.6 and 0.8.

As seen in the bottom panel of Figure 7, the objects with higher RB score generally are from the magnitude range 16 −19.5, as those form a significant fraction of the training set for the RB classifier.

### 3.2.2. Estimating Completeness Using Known Asteroids

The image differencing software can not distinguish between genuine transients and asteroids. We can take advantage of this to estimate completeness of detection by listing all asteroids expected to be present in a given field at a given time. By controlling which subset we look for, we can limit the expected error in position. Additional factors that need to be considered are gaps between CCDs, and pre-filtering cuts mentioned in Section 2 (e.g., detections within 10 pixels of CCD edges are ignored). We continue to investigate different fields to understand the possible loss of real objects.

### 3.2.3. Analysis of False Positives and Missed Detections

Throughout the survey lifetime, marshal users will continue to identify specific examples of false positives and missed detections that may have passed routine filters. The treeinterpreter module[44] can be used to provide insight into why certain transients receive a particular RB score. This module analyzes the trained scikit-learn classification model and breaks down the RB score of an individual transient into the sum of the contributions of each feature and a bias term. It can therefore be used to identify problematic features for certain kinds of transients or artifacts. This may then inform decisions over how features are calculated or which features should be included in future versions of the classifier.

### 3.3. Planned Deep Learning Implementation

The random forest based model described above is an improved version of a similar model used with PTF and iPTF. While random forest is effective, it still relies on features extracted by the team based on domain knowledge and sound mathematical and statistical concepts. Care is taken to reduce redundancies and irrelevancies that may have crept in as one starts with a large set of features. The improving success with better training sets stands witness to the efficacy. However, it is certainly possible that we miss out on some complex combinatorial features with significant discriminatory power to separate the real objects from the bogus ones. With this in mind, we have started exploring the use of convolutional neural networks (CNNs) trained on the science, reference, and difference image triplets.

We have used Xception CNN—where Inception modules are replaced with depthwise separable convolutions (Chollet 2016) —for our initial experiments. We initially experimented with PTF images and streaks and then moved on to the ZTF triplets. Using a Tesla K40 GPU for the CNN we find that both the FPR and FNR are better than for random forest. The downside is that the processing time needed for the CNN is an order of magnitude more than for random forest on a multicore CPU, and postage stamps must be made available to the classifier in real time. Since CNNs need more training data, we expect to continue experimenting over the next year and then adopt the CNN model. Another ongoing experiment involves using just the science and reference images as input during training to distinguish between real and bogus events, thereby reducing the processing time. In the future this will likely be combined with the generation of a salience map, and direct detection of transients based on the science and reference images using an encoder-decoder network (Sedaghat & Mahabal 2018).

---
[44] https://github.com/andosa/treeinterpreter/

Table 3
Features of the Current Streak RF Model

| Parameter | Description |
| --- | --- |
| flux | Integrated flux of the streak |
| bg | Local background level |
| length | Length of the streak |
| sigma | Sigma level of the streak |
| lengther | Error of the length of the streak |
| sigmaerr | Error of the sigma level of the streak |
| paerr | Error of the positional angle of the streak |
| bgerr | Error of the local background level |
| fitmagerr | Error of the fit magnitude |
| apsnr | S/N of the flux within the aperture |
| apmagerr | Error of the aperture magnitude |
| dmag | aperture magnitude - fit magnitude |
| dmagerr | Error of dmag |
| chi2 | $\chi^2$ of the fit |
| numfit | Number of attempts for a converged fit |

## 4. Solar System Objects

ZTF has a separate solar system pipeline looking for moving objects. We describe here ML aspects of some related initiatives.

### 4.1. Detection of Fast Moving Objects

Discovery of Near-Earth objects (NEOs) is one of ZTF's science goals. Fast moving objects (FMOs) are NEOs passing close to Earth moving at an angular rate higher than a few degrees per day. This causes their images to be smeared/trailed in a typical survey exposure (e.g., 30s), presenting a challenge for conventional moving object detection algorithms. To support ZTF's NEO activity, we have optimized a pipeline to detect trailed objects based on a PTF/iPTF prototype (Waszczak et al. 2017).

The pipeline starts by identifying contiguous pixels ("streaks") in the differenced ZTF images (Laher et al. 2014) similar to its PTF prototype. A trail-fitting model is used to analyze the streaks by computing their morphological features (Vereš et al. 2012). These measurements are used to train a random forest classifier, with the current set of features tabulated in Table 3. We currently rely on synthetically generated "reals" for training the model because the number of true FMOs is small (a few per week) and biased toward brighter objects.

The current training sample includes 50,000 synthetic trailed FMOs (with integrated AB magnitude brighter than 18.5) and an equal number of bogus streaks (extracted from real observations, see Figure 8). We tentatively adopt a threshold of 0.1 for the ML score (where 0 refers to definitely *bogus* and 1 to undoubtedly *real*) though streaks with a score lower than 0.1 still get inspected on a best-effort basis. Based on the examination of NEOs supposed to trail in ZTF images, we estimate a completeness of ∼70% for FMOs brighter than magnitude of 18.5. Since the beginning of operation in early 2018 February, this effort has led to a discovery of five FMOs (Bellm et al. 2019; Ye 2018). Current efforts are aimed toward accumulating known trailed FMOs for testing and verification, constructing a more realistic synthetic generation model, and improving the efficiency on the fainter end.

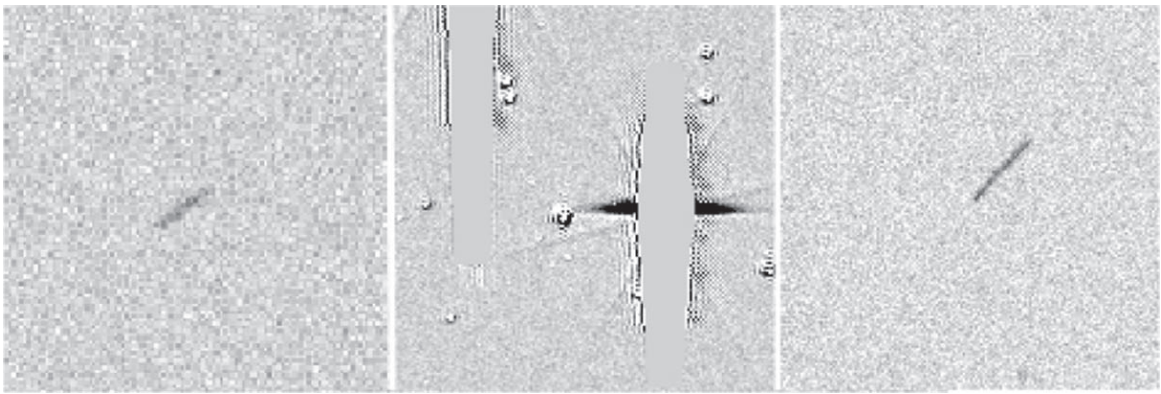

**Figure 8.** Typical example of a synthetic streak (left), a bogus streak (center), and a real FMO (right).

Given that the streaks left by FMOs are visually distinct, they provide a good classification opportunity for image-based convolutional neural networks (CNNs). We are experimenting transfer learning with certain canned networks e.g., ResNet (He et al. 2015), and also implementing some from scratch to ensure that we avoid overlearning. The inputs to the network are difference images. As the reference images do not contain the FMO, the difference image tends to include a clean streak. The contaminants include cosmic rays, airplane trails, streaks from bright stars, as well as some ghosting patterns. We have started experimenting with a pytorch[45] implementation. Because there are ways to eliminate some types of the contaminants upstream based on the random forest and CNN models, that is where the current efforts have concentrated. We use a network architecture with two convolutional layers with rectified linear units (ReLU) and a dropout fraction. For the two class problem we get an accuracy of 97.6%, and a FPR of 1.5% on a sample of ∼600 objects. As we start getting purer samples we will be implementing a real-time CNN module to mark FMO candidates for human inspection.

### 4.2. Determination of Asteroid Rotation Periods

Asteroid brightness ($V$) depends on four factors: (1) absolute magnitude ($H$); (2) heliocentric ($r$) and geocentric ($\triangle$) distances; (3) rotation ($\delta$); and (4) phase function ($\phi(\alpha)$), and can be written as

$$V = H + 5\log_{10}(r\triangle) + \delta - 2.5\log_{10}[\phi(\alpha)]. \quad (1)$$

In general, $\delta$ can be approximated by a second-order Fourier series to measure asteroid rotation period, and traditionally, the reliability of this periodic analysis is determined by human inspection of the folded light curve.

With ZTF, we expect to obtain hundreds of thousands of asteroid light curves every year. Because, after the period-finding step, inspection by humans of this huge data set is challenging, we have integrated a random forest algorithm into our periodicity analysis. The training samples are obtained from iPTF (Chang et al. 2014, 2015; Waszczak et al. 2015), and seven parameters extracted from the periodograms and the light curves (see Table 4) are used for machine learning. At this stage, we achieve a true-positive rate of ∼80% and a false positive rate of under 10%. This is based on an unbalanced sample of about 10,000 in the validation set with ∼15% having good periods.

### 5. Ongoing and Near-future Plans

The emphasis so far has been to understand the CCD characteristics, possible cross-talk, extent of streaking due to saturation etc. in order to understand non-astrophysical signatures (*bogus*) and separate them from the stream. Every clear night ZTF generates about 1 TB of images per night, and ∼$10^6$ alerts ($5\sigma$). Different science teams have been applying different filters to the remaining genuine candidates (*reals*) in near-real time with the aim of obtaining follow-up to understand their nature. The alerts contain a short history of each source. As time passes. longer light curves are starting to accumulate which can provide additional information for the sources. As light curves are available for all sources i.e., not just the ones that passed the real-time alert generation criteria, we can do statistically significant studies targeting variable sources as well. Different representations of the light curves, and combining ZTF data with other surveys opens up additional doors. As this is early in the life of ZTF we do not yet have enough uniform data from the survey itself for statistical studies. However, we outline some planned studies so that others can do parallel or complimentary projects using the public MSIP data.

---
[45] http://pytorch.org/

Table 4
Parameters used in the Random Forest Algorithm for Asteroid Rotation Period

| Parameter | Description |
| --- | --- |
| Base level of periodogram | The mean value $\chi^2$ of lowest 84% in the periodogram |
| Variance of periodogram | The standard deviation of $\chi^2$ of lowest 84% in the periodogram |
| Significance of nominated frequency | $|\chi^2$ of nominated frequency - base level of periodogram $|$ / variance of periodogram |
| Amplitude | 90% range of the light curve |
| Median mag error | Median error of magnitude in the light curve |
| Mean mag | Median magnitude in the light curve |
| Number of detections | Number of detections in the light curve |

### 5.1. Classification of Periodic Variables

We will use the ZTF light curves of variable stars to classify them using a method similar to Richards et al. (2011). As input to the machine learning classifier we will use (a) light curve statistics (e.g., Richards et al. 2011; Nun et al. 2015; Kim & Bailer-Jones 2016; Sokolovsky et al. 2017), (b) the output from period-finding algorithms like Lomb-Scargle periodogram (Lomb 1976; Scargle 1982), analysis of variance (Schwarzenberg-Czerny 1989), conditional entropy (Graham et al. 2013) and boxed least squares (Kovács et al. 2002), and (c) the fourier decomposition of the folded light curves. In addition, we will match the ZTF stars to external catalogs, including PS1 DR1 (Chambers et al. 2016), which adds additional color information, and the Gaia DR2 catalog (Gaia Collaboration et al. 2018a), which adds distance information.

To build a training set we will use catalogs of known variable objects (e.g., Drake et al. 2014; Gaia Collaboration et al. 2018b). We will use outlier detection methods to eliminate (a) erroneous identifications, (b) samples that have bad ZTF photometry, and (c) objects that are otherwise unsuited for the training sample. As such a use of external catalogs can introduce selection bias, we will also create an independent manually labeled test sample, from randomly chosen ZTF variable objects, to verify the classifier performance.

We will initially limit our classification effort to random forest (Breiman 2001), training the classifier to identify the major variability classes. An active learning strategy will be used to improve the classifier (e.g., Richards et al. 2012). In addition, we will compare our identification with catalogs of automatically identified variables by other telescopes, e.g., ATLAS (Heinze et al. 2018), Gaia (e.g., Mowlavi et al. 2018; Clementini et al. 2018), and ASASSN (Jayasinghe et al. 2018). Objects for which the classifications differ will be inspected and added to the training sample with it's correct label. After we have created a sizable training sample, we will experiment with more sophisticated classifiers like stochastic gradient boosting (Chen & Guestrin 2016) and neural networks (e.g., LeCun et al. 2015) which have been shown to perform better than random forest (Pashchenko et al. 2018).

### 5.2. Alternate Representation of Light Curves

Deep learning is finding its way to more and more applications. With ZTF we have several avenues to apply this cutting edge technique. An example is the *dmdt* method (Mahabal et al. 2017) where a light curve is cast into a 2D image by considering time and brightness difference between each pair of points. This density plot is like a structure function and can be used to differentiate between different classes. It can also be used as an initial step for transfer learning to identify new objects belonging to a class based on shorter light curves. Long short-term memory (LSTM) and Recurrent Neural Network (RNN) based light curves are another classification tool that can take advantage of the patterns in light curves despite their irregularity (see e.g., Naul et al. 2018; Charnock & Moss 2017). Such techniques will especially be useful for the irregular length ZTF light curves.

### 5.3. Domain Adaptation Using Other Surveys

Accumulating long enough light curves to extract unambiguous features can take months for a survey. One path to early characterization is through domain adaptation (DA) by taking advantage of past observations of the same sources, or other sources of similar classes from earlier surveys like CRTS and PTF. In short, if we have two classes of separable objects in some feature space of a Source survey (S), we can define a hyperplane to separate the two types. In a second Target survey (T), for the same features the hyperplane would be inclined differently. DA methods get the mapping between the two hyperplanes using a small fraction of data from the Target (T) survey and can then be used to predict the classes of the remaining majority of data in T. For the shorter ZTF light curves that we will have access to in the early days, being able to use existing labeled information to conduct classification can be cost-effective.

## 6. Synergistic Efforts with Brokers

While astronomers have been making good headway in handling small sets of objects, that is, in some sense, the status quo. ZTF allows astronomers to scale sample sizes by orders of magnitudes. The large sets of objects can be overwhelming for individuals, but that is exactly where computing resources combined with automated methods can be a boon. The scalable Kafka system is capable of handling the nightly hundreds of thousands of alerts. A string of brokers is expected to ingest and annotate the alert stream such that individuals can then query the brokers for smaller, pertinent sets of objects much like how the transient marshal currently does with rules defined by science teams.

ZTF time is split as public (MSIP, 40%), partnership (ZTF collaboration, 40%), and private (Caltech, 20%). Certain initiatives within the collaboration will handle the partnership alerts whereas the public brokers will cater to the MSIP portion of alerts. Given below are some brokering efforts in development that are likely to be associated with ZTF.

*AMPEL.* Photometric classification of transients will be carried out as part of the Alert Management, Photometry and Evaluation of Lightcurves (AMPEL) framework.[46] AMPEL is a public tool for alert analysis, where users create *channels* through the specification of filter criteria, analysis modules and triggered reactions. This can be easily done through the use of preexisting modules or made arbitrarily advanced as channels can include custom, user provided analysis algorithms implemented through a common python interface. All channels are integrated into the AMPEL system and consistently applied to the full ZTF alert stream. Alerts that are accepted are added to the live transient database for continued monitoring, reaction and follow-up. The live AMPEL instance is complemented by archived previous software configurations as well as archived alerts. These allow seamless data exploration through "mixing" software versions with alert streams.

*ASTROstream.* Automated claSsification of Transient astRonomical phenOmena (ASTRO) of ZTF alerts stream is an analysis pipeline based on the Lambda Architecture[47] (LA; Marz & Warren 2015) and using Apache Spark Streaming.[48] LA, a scalable and fault-tolerant data processing architecture, is designed to handle both real-time and historically aggregated batched data in an integrated fashion. Spark is a cluster computing framework which is widely used as an industry tool to deal with big data processing and contains built-in modules for streaming and machine learning (Peloton et al. 2018, see also Huang et al. 2017). LA enables a continuous processing of real-time observation via speed layer. This layer ingests streaming alert as it is generated and analyzes data in real time to get insight immediately and provides potential targets for follow-up to space- and ground-based telescopes.

The Batch transient classier engine (Batch layer) is based on a deep learning architecture. It ingests large batches of data, with counterparts cross-matched from the PS1, SDSS, and other catalogs, in order to extract the best features to classify transients in the data set (Golkhou et al. 2018). Similar in concept to transfer learning, extracted features by the Batch engine (see e.g., Richards et al. 2011; Faraway et al. 2016) can update the feature space of the Real-time engine (i.e., replace the model/classifier in the speed layer with the trained model in the Batch layer.)

*Antares.* The Arizona-NOAO Temporal Analysis and Response to Events System (ANTARES) is a broker that has resulted from a collaboration between NOAO and University of Arizona (Saha et al. 2014). Expected to work on sparse, unevenly sampled, heteroskedastic data, it is built with goals of early, intermediate and retrospective classification of alerts to cater, respectively, to early categorization, multi-class typing, and a purity-driven subtyping of the stream events (Narayan et al. 2018).

*Lasair.* Based in the United Kingdom, and focused on the community there, Lasair is a broker for rapid transients. Currently funded, and built upon existing work on following up transients and carrying out machine learning[49], the group is aiming to be a LSST Community Broker. It will provide cross-matches, cutouts, probabilistic classifications, and will have API access, and specific dataflows from researchers will also be supported.

*ALeRCE.* Out of Chile and incorporating US scientists, Automatic Learning for the Rapid Classification of Events (ALeRCE[50]), combines astronomy, machine learning and statistics and focuses on real-time classification of non-moving objects. It is based on a federated and hierarchical classification model, with multiple classifiers specialized on different variability classes downstream from a central node. ALeRCE will provide an annotated stream, including cross-matches and class probabilities, as well as visualization tools for the analysis of individual alerts and sections of the stream.

*MARS.* The Las Cumbres Observatory (LCO) has made available a set of tools for public access of ZTF alerts. Called *Making Alerts Really Simple*, or MARS[51], this is not a full broker in that it does not have characterization built in, and just provides access to the alert stream along with many ways to subselect sources. This enables many downstream activities.

The fast multiplying data volume implies an opening up for data driven methodologies. For scalability, methods deployed on clusters, GPUs, and other multi-processing hardware (e.g., Gieseke et al. 2015, 2017) are expected to find increasing use in all the brokers. Methods to save follow-up time through, for instance, optimizing number of spectra needed will also find greater use (e.g., Ishida et al. 2019; Vilalta et al. 2017) as astronomers subscribe to the brokers and pick their rivulet of objects.

## 7. Conclusions

ZTF is well underway and set to produce a remarkable number of transients and variable observations. We are starting to exercise various machine learning techniques starting with real/bogus and star-galaxy separation, and leading to early detection and classification of moving and non-moving objects. Another set of techniques is being deployed to enable archival studies and domain adaptation. Once the public brokers come online the usage by the community is expected to explode. Given that any ZTF source can be studied spectroscopically with a 1–5$m$ class telescope, large studies will be possible for many families of objects as well as outliers. In a way, ZTF is a precursor to the much larger LSST survey (see Table 1 in Graham et al. 2019, for a comparison between ZTF and LSST). Moreover, LSST will be able to treat the ZTF data set as a stepping stone for the domain adaptation process, enabling a quick start to many different analyses. We are now truly entering a real-time era of large transient data sets.

---

[46] ampelproject.github.io/Ampel/
[47] http://lambda-architecture.net/
[48] http://spark.apache.org/streaming/
[49] https://star.pst.qub.ac.uk/
[50] http://alerce.science/
[51] http://Mars.lco.global/

Based on observations obtained with the Samuel Oschin Telescope 48-inch and the 60-inch Telescope at the Palomar Observatory as part of the Zwicky Transient Facility project. Major funding has been provided by the U.S. National Science Foundation under Grant No. AST-1440341 and by the ZTF partner institutions: the California Institute of Technology, the Oskar Klein Centre, the Weizmann Institute of Science, the University of Maryland, the University of Washington, Deutsches Elektronen-Synchrotron, the University of Wisconsin-Milwaukee, and the TANGO Program of the University System of Taiwan. Part of this research was carried out at the Jet Propulsion Laboratory, California Institute of Technology, under a contract with the National Aeronautics and Space Administration.

*Facilities:* PO:1.2m, PO:1.5m.

## ORCID iDs

Ashish Mahabal  https://orcid.org/0000-0003-2242-0244